\documentclass[aps,twocolumn,showpacs,prb,amsmath,amssymb,superscriptaddress]{revtex4-1}
\usepackage{color}
\usepackage{amsmath}
\usepackage{verbatim}
\usepackage{hyperref}
\usepackage{epsf}
\usepackage{graphicx}
\usepackage{dcolumn}
\usepackage{bm}

\begin{document}
\title{ Microscopic theory of magnetism in Sr$_3$Ir$_2$O$_7$}
\author{Jean-Michel Carter}
\affiliation{Department of Physics, University of Toronto, Toronto, Ontario M5S 1A7 Canada} 
\author{Hae-Young Kee}
\email{hykee@physics.utoronto.ca}
\affiliation{Department of Physics, University of Toronto, Toronto,
Ontario M5S 1A7 Canada}
\affiliation{Canadian Institute for Advanced Research, Toronto, Ontario  Canada}

\begin{abstract} 
	
	An intriguing idea of spin-orbit Mott insulator has been proposed to explain magnetic insulating behavior in various iridates. 
	This scenario relies on the strength of the spin-orbit coupling being comparable to electronic correlations, and it is not {\it a priori} obvious
	whether this picture is valid for all iridates.  In particular, Sr$_3$Ir$_2$O$_7$ exhibits a small charge gap and magnetic moment
	compared to Sr$_2$IrO$_4$, questioning the validity of such hypothesis.
	To understand the microscopic mechanism for magnetism in Sr$_3$Ir$_2$O$_7$, we construct a tight binding model
	taking into account the full $t_{2g}$-orbitals, the staggered rotation of the local octahedra, and the bilayer
	structure. The bands near the Fermi level are mainly characterized by the total angular momentum
	$J_{\text{eff}}=1/2$, except below the $\Gamma$ point, supporting a reasonably strong spin-orbit coupling picture.
	A first order transition to a collinear antiferromagnet via multi-orbital Hubbard interactions
	is found within the mean field approximation.  The magnetic moment jump at the transition is consistently smaller than
	Sr$_2$IrO$_4$, originated from the underlying band structure of a barely band insulator. 
	Given the small charge gap and moment observed in Sr$_3$Ir$_2$O$_7$, the system is close to a magnetic transition.
	A comparison to a spin-model is presented and connection to the Mott insulator is also discussed.

\end{abstract}

\maketitle

\section{Introduction}
The concept of Mott insulators is prevalent in strongly correlated materials, referring to a strong electronic correlation driven insulator that
violates the conventional band theory. Often, Mott insulators are accompanied by antiferromagnetic ordering, but genuine Mott insulators have a robust charge gap above the N\'{e}el temperature  where the magnetic ordering disappears. 
Mott insulators have been found in correlated electronic systems with 3d-orbital materials including the high transition temperature cuprates, and
are considered seeds of exotic phases when doped by holes or electrons.
However, with heavier atom systems, the outer shell electron wavefunctions are less localized leading to
weaker electronic correlation, and thus band theory should be a good starting point
to model the behavior of electrons in such solids. 

Surprisingly, Sr$_2$IrO$_4$, a material with heavy 5d Ir atoms, exhibits a magnetic insulating state, despite having partially filled 5d-orbitals.
One missing ingredient that is relevant in these heavy elements is spin-orbit coupling (SOC).
In particular, when the SOC strength is comparable to that of electronic interactions,
understanding their interplay becomes challenging.
Iridates with 5d-orbitals offer such a playground to investigate their combined effects.
%
It has been suggested that due to the narrowing of the bandwidth induced by the strong SOC,
the effect of Hubbard interactions is amplified, leading to an insulating state in some layered
perovskite\cite{Crawford:1994ys,Cao:1998uq,Cao:2002uq,Nagai:2007vn,Kim:2008ve,Moon:2008ly,Jackeli:2009qf,Kim:2009bh,Fujiyama:2012}
and pyrochlore\cite{PesinBalents:2010,Yang:2010uq} iridates.
This state was called a spin-orbit Mott insulator.\cite{Kim:2008ve,Moon:2008ly,Kim:2009bh}
%
While this proposal of Mott insulator for Sr$_2$IrO$_4$ itself was challenged in recent studies\cite{Arita:PRL, Hsieh:PRB},
the idea has been quickly applied to other iridates including Sr$_3$Ir$_2$O$_7$\cite{BJKIM2,Wang:arXiv1210.4141}, its bilayer sister.

However, optical conductivity\cite{Moon:2008ly} and transport\cite{Ge:2011yq,Cao:2002uq} data have shown that the charge gap in the bilayer system is significantly smaller than in the single layer compound. In addition, the magnetic structures in these two systems differ; Sr$_3$Ir$_2$O$_7$ displaying a collinear antiferromagnetic structure with moments aligned with the c-axis\cite{BJKIM:2012_arxiv, BJKIM:2012_arxiv2, Dhital:100401, Boseggia:2012bb, Clancy:2012_arXiv2} whereas Sr$_2$IrO$_4$ is well-known to have a canted antiferromagnetic structure with moments in the ab-plane\cite{Kim:2008ve, Kim:2009bh, Jackeli:2009qf}. This may be related to the crystal structure of the two compounds. Sr$_3$Ir$_2$O$_7$ crystallises in a Bbcb space group\cite{Matsuhata20043776} whereas Sr$_2$IrO$_4$ does with an I4$_1$/acd space group\cite{Crawford:1994ys}. Another major difference between these two materials is the giant magnon gap seen in Sr$_3$Ir$_2$O$_7$\cite{BJKIM2} whereas in Sr$_2$IrO$_4$, the magnon gap, if it exists, is too small for detection with current resonant inelastic x-ray scattering (RIXS) resolution\cite{JunghoKim}. 
Based on these differences, it is not clear whether a strong SOC approach is valid Sr$_3$Ir$_2$O$_7$.

In this paper, we build a tight-binding Hamiltonian for the bilayer Sr$_3$Ir$_2$O$_7$ taking into account the full $t_{2g}$ manifold, the local staggered rotation of the octahedra, and
bilayer coupling, to understand the validity of strong SOC picture and the interplay among SOC, electronic correlation, and the crystal structure.
The paper is organized as follows.
In Sec. \ref{sec:tbmodel}, using a Slater-Koster theory\cite{Slater:1954tg}, the tight binding parameters are estimated. Comparing this band structure
with recent angle-resolved photoemission spectroscopy (ARPES)\cite{Wang:arXiv1210.4141, Wojek:JPCM2012} measurements and first-principle calculations\cite{Moon:2008ly}, we determined the SOC strength.
Indeed the bands near the Fermi level are mainly $J_{\text{eff}}=1/2$, except near the $\Gamma$ point below the Fermi level, favoring a reasonably strong SOC limit. Due to the nature of $J_{\text{eff}}=1/2$ wavefunction, Sr$_3$Ir$_2$O$_7$ bilayer material exhibits a nearly band insulator {\it distinctly} different from Sr$_2$IrO$_4$.
This originates from large bilayer hopping terms
and the alternating rotations of the local octahedra.
We then study the effects of electronic interactions using a multi-orbital Hubbard model with Hund's coupling in Sec. \ref{sec:mftheory}.
A first order transition is found within mean field theory, and the jump in the magnetization and the critical interaction strength depend on the SOC value. 
The implications of our study in the context of recent experimental results are discussed in the section \ref{sec:discussion}.

\section{Tight-binding model}
\label{sec:tbmodel}
%
\begin{figure}[!h]
\includegraphics[width=3.5in,angle=0]{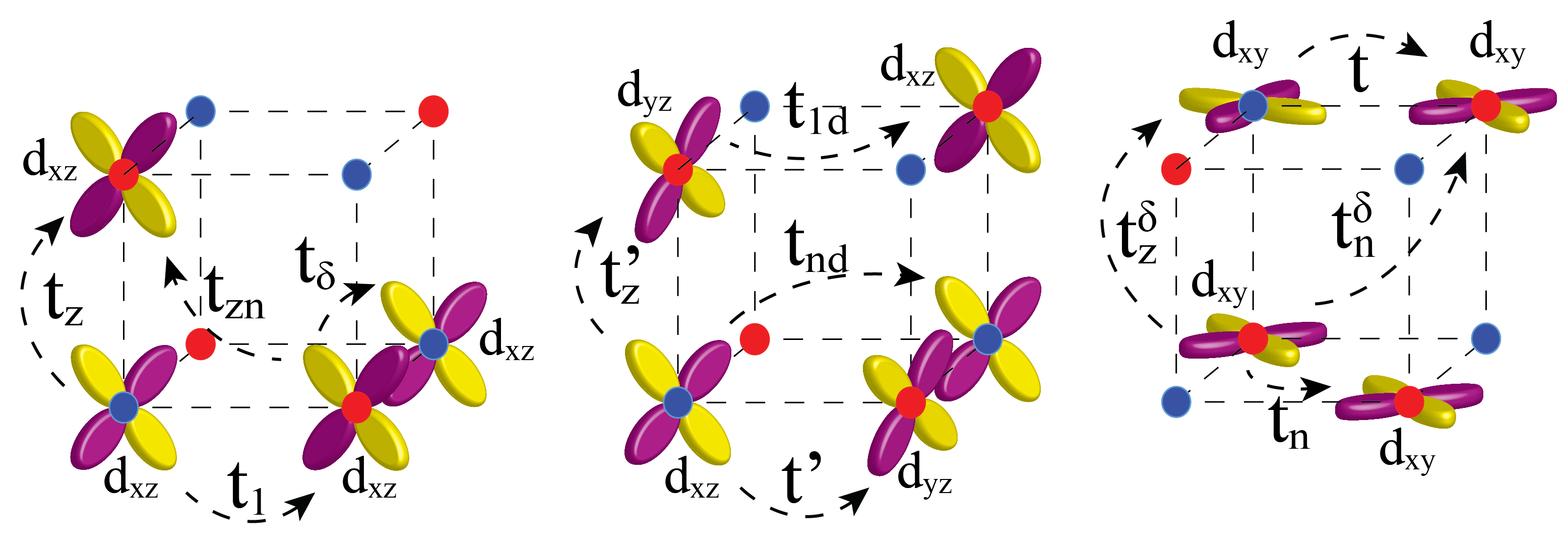}
\caption{[color online] Tight binding hopping parameters: the red and blue dots represent two different Ir atom environments due to 
staggered octahedra rotation. $(t,t_1,t_z,t_\delta,t_z^\delta)$ are NN hopping between the same orbitals, while $(t',t'_z)$ between different orbitals.
$(t_n,t_{zn},t_{nd},t_n^\delta)/(t_{1d})$ are next NN hopping integrals between the same/different orbitals.}
\label{Fig:Hoppings}
\end{figure}

The bilayer structure is represented by a dashed cube in Fig. \ref{Fig:Hoppings}, with the red and blue vertices denoting the two different Ir atoms in the bilayer structure due to IrO$_6$ staggered rotation\cite{Subramanian:1994fk} (the in-plane unit cell area is thus doubled).
Taking into account the $t_{2g}$ d-orbital $\alpha = yz,xz,xy$ with momentum ${\bf k}$, sublattice $\gamma$ = blue(B), red(R),  
and spin $\sigma$, the tight binding Hamiltonian can be written as
 $\sum_k\psi_{k, l}^\dagger H^{ll'}_{0} \psi_{k,l'}$, with spinor $\psi_{k,l} = (d_{k,\downarrow}^{l,B,yz}, d_{k,\downarrow}^{l,B,xz}, d_{k,\uparrow}^{l,B,xy}, B \Leftrightarrow R , \uparrow \Leftrightarrow \downarrow)^T$
where $d_{k,\sigma}^{l,\gamma,\alpha}$ is a annihilation operator at the layer $l,l'=1,2$, sublattice $\gamma$ and orbital $\alpha$. 
Using this basis, the matrices $H_{0}^{ll'}$ including nearest neighbor (NN) and next NN hoppings, is given by
\begin{equation}
H^{ll'}_{0} =
\begin{pmatrix}
	H_{so} \delta_{ll'} + H^{ll'}_{BB}	& H^{ll'}_{BR} \\		
	H_{BR}^{ll' \dagger}	& H_{so} \delta_{ll'}  + H^{ll'}_{RR} 
\end{pmatrix}
	+ [\text{time-reversed}],
	\label{eq:SLHamil_t2g}
\end{equation}
with $H_{so}$ being the atomic spin-orbit coupling, $\lambda {\bf L}_i \cdot {\bf S}_i$.
The intra-layer intra-sublattice hopping (next NN) of $H^{ll}_{BB} = H^{ll}_{RR}$ are given by
\begin{equation}
H_{so}  + H^{ll}_{BB} = 
\begin{pmatrix}
	\epsilon^{yz} & -i \frac{\lambda}{2} +\epsilon_{1d} & \frac{\lambda}{2} \\
	i \frac{\lambda}{2}+\epsilon_{1d} & \epsilon^{xz} & i \frac{\lambda}{2}\\
	\frac{\lambda}{2} & -i \frac{\lambda}{2} & \epsilon^{xy}
\end{pmatrix},
	\label{eq:Hso}
\end{equation}
where 
$\epsilon^{xy} = 4 t_n \cos(k_x)\cos(k_y) + \mu_{xy}$, $\epsilon^{yz} =\epsilon^{xz}= 4 t_{nd} \cos{k_x} \cos{k_y}$, and $\epsilon_{1d} = 4 t_{1d} \sin{k_x} \sin{k_y}$.
$\mu_{xy}$ denotes the atomic potential difference between $xy$ and one-dimensional $xz/yz$ orbitals due to tetragonal distortion. 
$H^{ll}_{BR}$ is NN hopping terms between B and R sublattice sites, and is written as 
\begin{equation}
H^{ll}_{BR} = 
\begin{pmatrix}
	\epsilon_k^{yz}		& \epsilon_k^{rot}	& 0 \\
	-\epsilon_k^{rot}	& \epsilon_k^{xz} 	& 0 \\
	0 					& 0 				& \epsilon_k^{xy}
\end{pmatrix},
	\label{eq:H_BR}
\end{equation}
where the dispersions are given by
$\epsilon_k^{yz} = 2 (t_1\cos(k_y) + t^{\delta}\cos(k_x))$,
	$\epsilon_k^{xz} = 2 (t_1\cos(k_x) + t^{\delta} \cos(k_y))$,
	$\epsilon_k^{xy} = 2 t  (\cos(k_x) + \cos(k_y))$,
and	$\epsilon_k^{rot} = 2 t' (\cos(k_x) + \cos(k_y))$.
For the bilayer hopping terms, it is important to notice that a R (B) atom on one layer lies on top of a B (R) on the other layer.
Therefore, the bilayer hopping terms in $H_0^{12}$ are given by  
\begin{equation}
H_{BR}^{12} = 
\begin{pmatrix}
	t_z		& t'_z	& 0 \\
	-t'_z	& t_z 	& 0 \\
	0 		& 0 	& t_{z}^{\delta}
\end{pmatrix};
H_{BB}^{12} =  H_{RR}^{12}=
\begin{pmatrix}
	\epsilon_{d}^{yz}		& 0						& 0 \\
	0						& \epsilon_{d}^{xz} 	& 0 \\
	0 						& 0 					& \epsilon_d^{xy}
\end{pmatrix},
	\label{eq:H_bl}
\end{equation}
where $\epsilon_{d}^{yz} = 2 t_{zn}\cos{k_y}$, $\epsilon_{d}^{xz} = 2 t_{zn}\cos{k_x}$, and $\epsilon_d^{xy}= 4 t^{\delta}_n \cos{k_x} \cos{k_y}$.
The hopping parameters of ($t$, $t_1$, $t_z$, $t'$, $t'_z$, $t_n$, $t_{zn}$, $t_{1d}$, $t^{\delta}$, $t_z^{\delta}$, $t_{nd}$, $t_n^{\delta}$) are shown in Fig. \ref{Fig:Hoppings}.
%

%
Setting $t$ as a unit, there appears to be 11 other hopping parameters, but they are not all independent of each other. Note that
in an ideal octahedra, $\mu_{xy}=0$,  $t = t_1 = t_z$, $t_n=t_{zn}$, and $t^{\delta}=t_z^{\delta}$ based on cubic symmetry,
while $t'=t'_z=0$ without the staggered rotation of octahedra.  
Due to the IrO$_6$ rotation, the above relations break down and how they differ depends on the angle of the staggered rotation. 
Using Slater-Koster theory\cite{Slater:1954tg} with $t_{dd\sigma} : t_{dd\pi} : t_{dd\delta}= 3/2: -1 : 1/4$ \cite{OKAnderson:1973}, 
and taking into account a distance factor of 0.9 for bilayer terms and of 0.2 for the next NN for the exponential suppression of hopping parameters with distance, we found
($t$, $t_1$, $t_z$, $t'$, $t'_z$, $t_n$, $t_{zn}$, $t_{1d}$, $t^{\delta}$, $t_z^{\delta}$, $t_{nd}$, $t_n^{\delta}$)
= ($-1.0$, $-0.94$, $-0.8$, $0.15$, $0.36$, $0.16$, $0.2$, $0.11$, $0.27$, $0.15$, $0$, $0$) for a staggered rotation angle of $\pm$12$^\circ$.
 Once the ratio between $t_{dd\sigma}$, $t_{dd\pi}$, and $t_{dd\delta}$ is fixed, only the SOC $\lambda$ and the tetragonal splitting $\mu_{xy}$ 
remain independent parameters.

\begin{figure}[t]
\includegraphics[width=3.5in,angle=0]{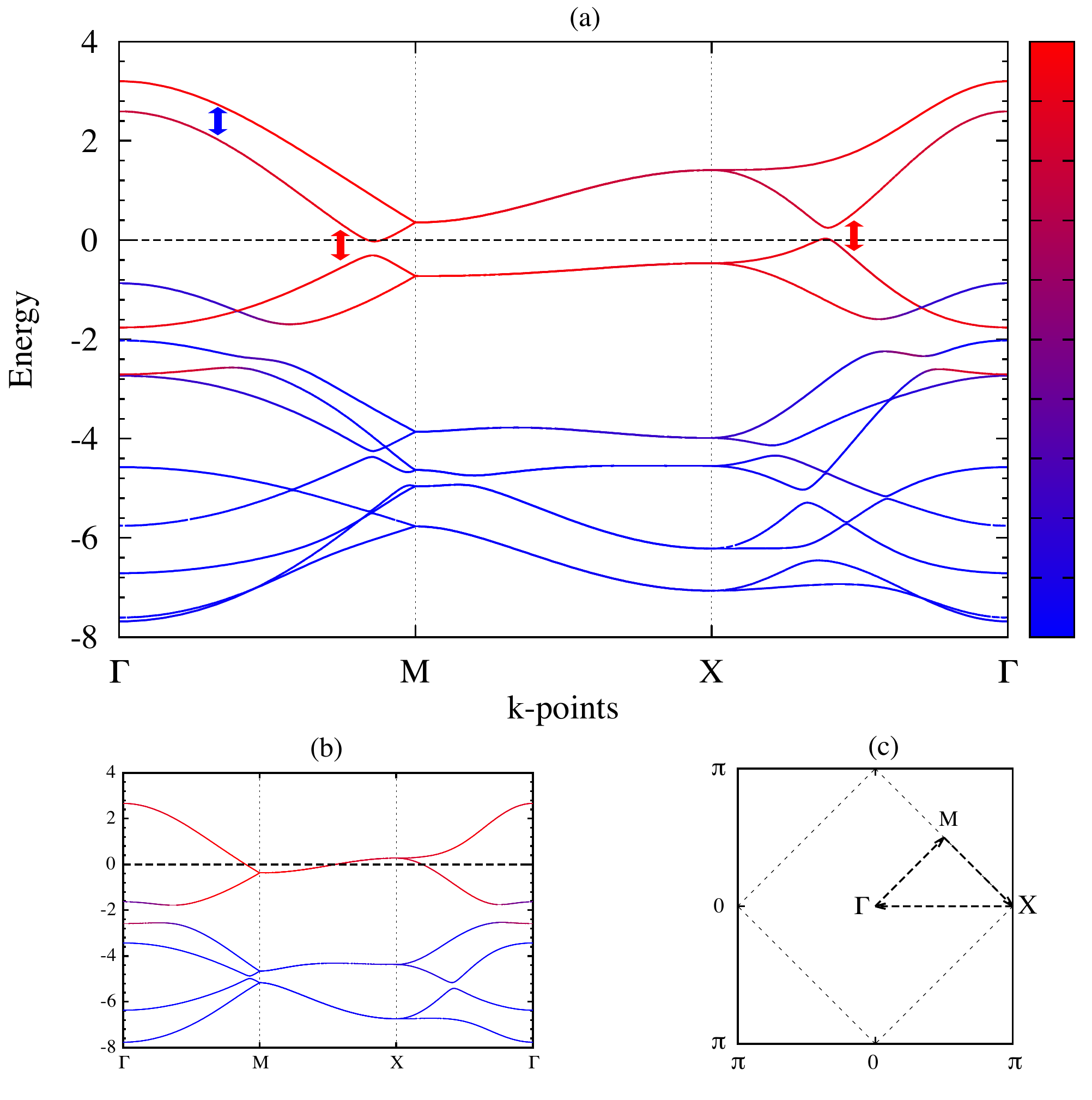}
\caption{[color online] (a) Underlying band structure of Sr$_3$Ir$_2$O$_7$ along the path shown in (c). The weight of the $J_{\text{eff}}=1/2$ state is colored in red while the $J_{\text{eff}}=3/2$ weight is in blue, and we see that the relevant bands near the Fermi level is mostly composed of $J_{\text{eff}}=1/2$ wavefunction, but note that the band immediately below the Fermi level around the $\Gamma$ point is made of $J_{\text{eff}}=3/2$ state. The band structure is distinct from the single layered Sr$_2$IrO$_4$ system shown in (b) mainly due to the large bilayer hoppings inherited from the nature of the $J_{\text{eff}}=1/2$ wavefunction.}
\label{Fig:bs}
\end{figure}

The tight binding band structure is shown in Fig. \ref{Fig:bs} for $\lambda/t = 3$ and $\mu_{xy}=0$, where
$J_{\text{eff}}=1/2$ bands are in red, while $J_{\text{eff}}=3/2$ bands are in blue. 
There are several important features to notice.
First,  $J_{\text{eff}}=1/2$ and $3/2$ bands below the Fermi level
are mixed along $\Gamma$ to $M=(\pi/2,\pi/2)$ and along $\Gamma$ to $X=(\pi,0)$, and the bands near $\Gamma$ are $J_{\text{eff}}=3/2$ states, not $1/2$.
Second, this band structure is similar to
recently reported ARPES data\cite{Wojek:JPCM2012,Wang:arXiv1210.4141} on Sr$_3$Ir$_2$O$_7$ and a first principle calculation\cite{Moon:2008ly}, 
but the bands near $\Gamma$ have been misidentified as $J_{\text{eff}}=1/2$
in these works.\cite{Wang:arXiv1210.4141, Moon:2008ly}
Third, our results imply that the SOC is large, but not enough to fully separate $J_{\text{eff}}=1/2$ and $3/2$ bands below the Fermi level,
similar to theoretical studies on Sr$_2$IrO$_4$\cite{Watanabe:2010cr,Arita:PRL}. On the other hand,
the unoccupied bands above the chemical potential are pure $J_{\text{eff}}=1/2$.  Thus the RIXS intensity\cite{}
contains more $J_{\text{eff}}=1/2$ band contribution, because it is from a combination of
unoccupied and occupied bands, while ARPES measures only occupied states. 

The $M-X$ path is the reduced Brillouin zone (BZ) boundary due to the staggered octahedra rotation, and the degeneracy is protected 
because the in-plane potential generated by the staggered rotation has the form of $\cos{k_x}+\cos{k_y}$ which is absent along this path.
The energy difference in $J_{\text{eff}}=1/2$ bands is largest at $X$ point, because of a constructive
combination of $t$, $t_z$, $t'$, and $t'_z$.
A typical splitting of $0.8t$ could be measured by the separation of the two unoccupied bands of $J_{\text{eff}}=1/2$ 
near the $\Gamma$ point (denoted by the blue arrow in Fig. \ref{Fig:bs}a).
In Ref. \onlinecite{Wang:arXiv1210.4141}, the bilayer splitting was estimated by the separation between the two highest occupied bands at $\Gamma$.
However, the energy difference between either the 1st and 2nd or the 1st and 3rd highest occupied band at the $\Gamma$ point in Fig. \ref{Fig:bs}(a)
is due to both the bilayer hoppings and the hybridization between $J_{\text{eff}}=1/2$ and $3/2$ bands.
This splitting further gives an estimate of our unit $t \sim 200$ meV, which is consistent with the overall bandwidth when comparing our tight binding spectrum with the ARPES data\cite{Wang:arXiv1210.4141}.

Last, the most important result is the topology of the band structure.  It is almost a band insulator where the Fermi level
barely touches the bottom and top of unoccupied and occupied bands near $M$ and $X$ respectively.
A small difference in the hopping parameters could change particular features
such as the size of the small hole and electron pockets,
but the separation of the two unoccupied bands from the rest via a direct gap at every $k$-point, does not depend on the details.
The direct gap between these two unoccupied bands occurs due to a finite $t'$ or $t'_z$, shown as the red arrow in Fig. \ref{Fig:bs}(a), indicating
the importance of staggered octahedra rotation between the NN Ir atoms. For comparison, we present the underlying band structure of the single layer iridates in Fig. \ref{Fig:bs}(b) that shows the large Fermi surface crossing along the M-X path.
%
%
Now that we are equipped with the proper tight binding model, let us move to a magnetic ordering mechanism.

\section{Mean-Field Theory}
\label{sec:mftheory}
A general interaction Hamiltonian in multi-orbital systems is given by
\begin{equation}
H_{int} = U \sum_i n_{i \alpha\uparrow}  n_{i \alpha \downarrow} + U'\sum_{i, \alpha \neq \beta} n_{i \alpha} 
n_{i \beta} - J \sum_{i \alpha \neq \beta} {\bf S}_{i \alpha} \cdot {\bf S}_{i \beta},
\end{equation}
where the Hund's coupling $J=(U-U')/2$ is determined by intra- and inter-orbital Hubbard $U$ and $U'$ and where we set $U' = 0.8 U$\cite{Mizokawa, Arita:PRL}.
We treat these interactions at the mean-field (MF) level in the magnetic channels to find possible magnetic
orderings.
To consider the strong SOC, we define the order parameter in the $J_{\text{eff}}$-basis. To do so, we first rewrite the above $H_{int}$ in the $J_{\text{eff}}$-basis,
and decouple all the terms in the magnetic channels. The mean field Hamiltonian is then given by
$H_{int}^{MF}=  - \left( \frac{U}{3} + \frac{2 U'}{3}-\frac{2 J}{3} \right) {\bf  m}_{i1} \cdot  {\bf S}_{i1} 
- \left( \frac{U}{2} + \frac{ U'}{2}-\frac{ J}{2} \right) \sum_{n=2,3} {\bf  m}_{in} \cdot  {\bf S}_{in}.$
Here ${\bf m}_{in} = \langle {\bf S}_{in} \rangle = \frac{1}{2}\langle c^\dagger_{in\mu} {\vec \sigma}_{\mu\nu} c_{in\nu} \rangle$, where
${\vec \sigma}$ is the Pauli matrix for pseudospin for $n=1,2,3$ 
that correspond to $(J_{\text{eff}},J_{\text{eff}}^z) = (1/2,\pm 1/2)$, $(3/2,\pm 3/2)$, and $(3/2,\pm 1/2)$,
respectively.
The order parameters $m_{i,n}$ are then determined self-consistently.
There are 36 MF order parameters from three pseudospin states ($n$), two sublattices ($\gamma$), two layers ($l$), and three directions of ${\bf m}$.

\begin{figure}[!ht]
\includegraphics[width=3.35in,angle=0]{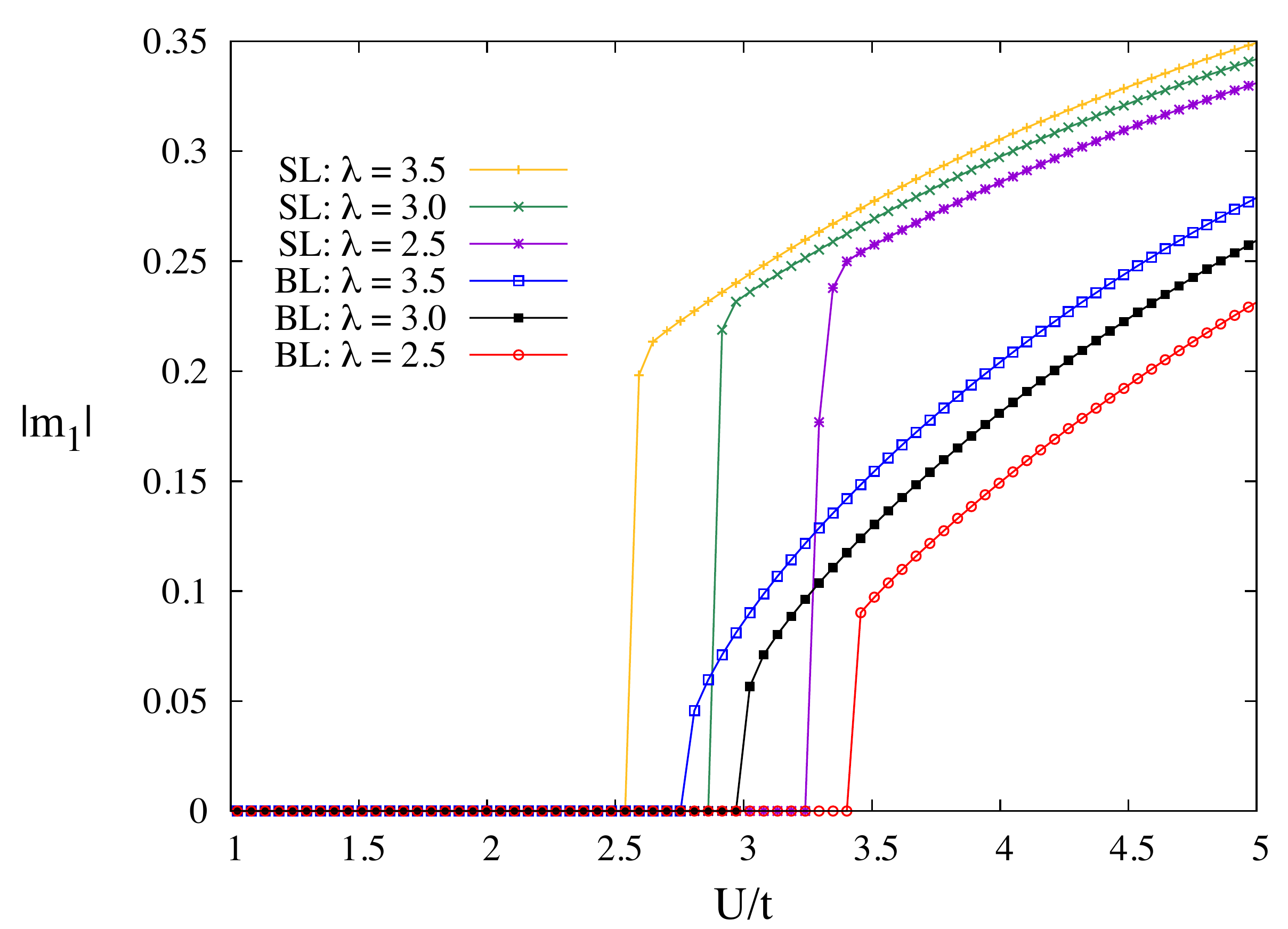}
\caption{[color online] First order phase transition is found for bilayer (BL) Sr$_3$Ir$_2$O$_7$ for $\lambda/t = $ 2.5, 3 and 3.5. For comparison, the transition for single layer (SL) Sr$_2$IrO$_4$ is also shown for the same SOC strengths. We note that for the same set of parameters, the Sr$_3$Ir$_2$O$_7$ requires a higher $U_c$ for the magnetic moment to set in, due to its underlying band insulator. Note that only {\bf $|$m$_1$$|$} is shown in the figure since {\bf $|$m$_2$$|$} and {\bf $|$m$_3$$|$} are negligible.}
\label{Fig:transition}
\end{figure}

\begin{figure}[!ht]
\includegraphics[width=3.35in,angle=0]{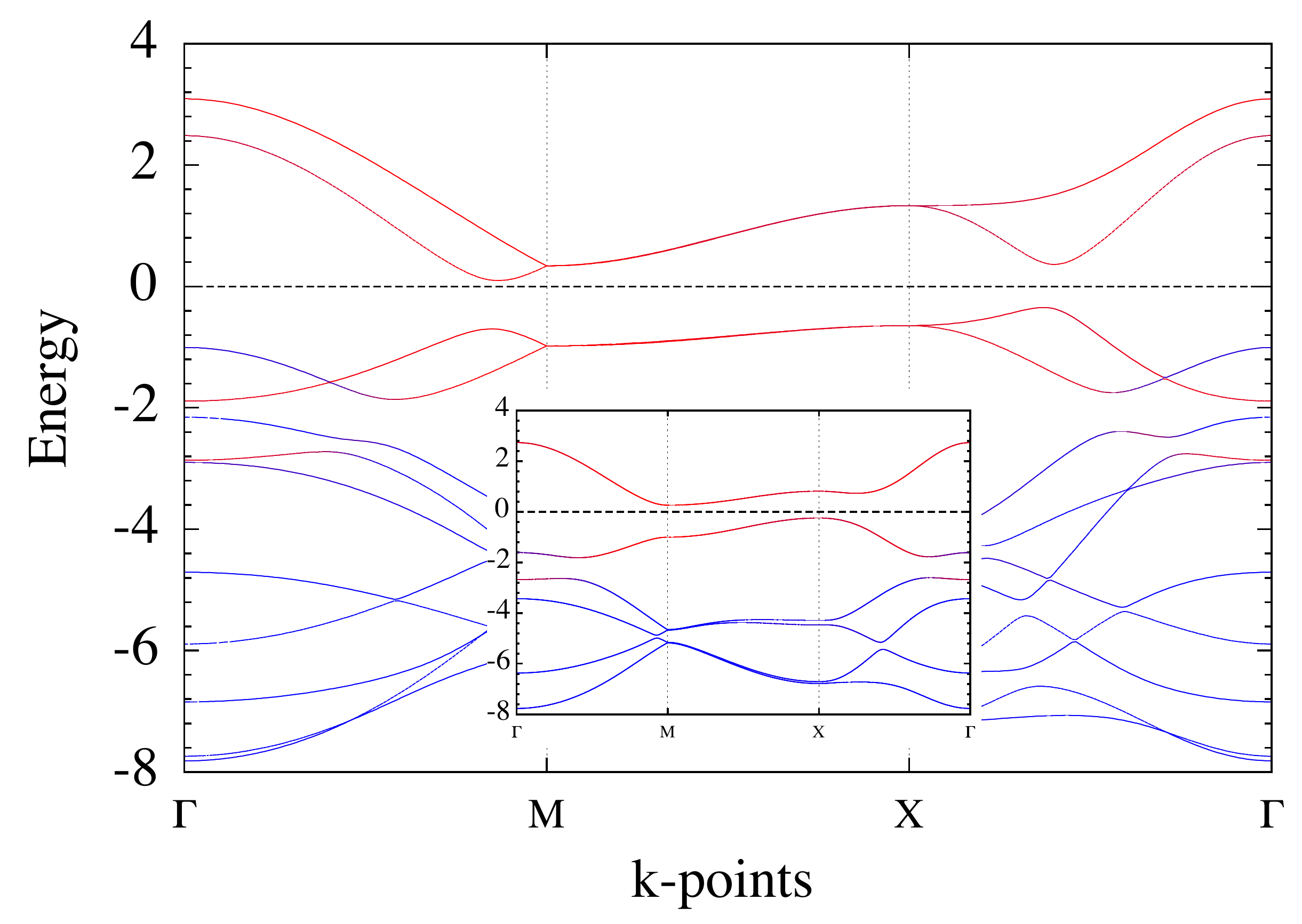}
\caption{[color online] Band structure with magnetic ordering for Sr$_3$Ir$_2$O$_7$ (and Sr$_2$IrO$_4$ in inset) using the self-consistent solution at $U/t = 4$. The charge gap from these calculation are $\Delta_c \sim 0.4t$ for Sr$_3$Ir$_2$O$_7$ and $\Delta_c \sim 0.6t$ for Sr$_2$IrO$_4$.}
\label{Fig:bs_ordered}
\end{figure}

As $U$ increases, there is a weakly first order phase transition from a barely band insulator into a magnetically ordered
phase as shown in Fig. \ref{Fig:transition}. ${\bf m}_1$ is finite and the magnetic pattern is
given by a G-type antiferromagnetic order where the moments in the blue and red atoms point in opposite directions. 
The ${\bf m}_2$ and ${\bf m}_3$ order parameters are smaller by two orders of magnitude, confirming that the moments are made of $J_{\text{eff}}=1/2$ electrons. The critical interaction $U_c$ depends on the SOC strength.
The bigger the SOC, the smaller the $U_c$ is, similar to the single layer case reported in Ref. \onlinecite{Watanabe:2010cr} and confirmed here,
but opposite to three dimensional materials, SrIrO$_3$\cite{Zeb:2012_arxiv}.


%
%

A finite moment affects the band structure, where the bottom of electron and top of hole bands are pushed away from the Fermi level as shown in Fig. \ref{Fig:bs_ordered}, making a charge gap visible. For comparison, we show the band structure after magnetic ordering for the Sr$_2$IrO$_4$ system also, where the bands are much more affected by the ordering than in the Sr$_3$Ir$_2$O$_7$ system.
When  $ U > U_c \sim 3t$ for $\lambda/t = 3$,  there are two competing ordering patterns -- canted antiferromagnet (AF) and collinear AF. 
We found that when the ratio of $t'_z/t_z$ is equal to the ratio of $t'/t$, the canted AF and collinear AF become degenerate.
From the Slater-Koster theory shown above,  $t'_z/t_z$ $\approx$ 2.4 $t'/t$, and a G-type collinear AF state along the
c-axis is favored by the bilayer coupling.
Such a state has been recently confirmed in experiments\cite{BJKIM:2012_arxiv, BJKIM:2012_arxiv2, Dhital:100401, Boseggia:2012bb, Clancy:2012_arXiv2}.
%

Given the small magnetic moment observed in Sr$_3$Ir$_2$O$_7$\cite{Fujiyama:2012}, the system is close to a magnetic transition, where the size of
the charge gap from transport data does not support a strongly correlated insulator.\cite{Cao:2002uq}
Indeed when the magnetic moment disappears, the band structure is almost
a band insulator where the Fermi level barely touches the bottom/top of electron/hole bands.
However, the magnetic moment appears only in $J_{\text{eff}}=1/2$ channel, and one may ask if this magnetic insulator is connected to the large $U$-limit of spin-orbit Mott insulator. The spin-model derived from a $J_{\text{eff}}=1/2$ only model\cite{JM_arxiv_2012_iridates} indeed displays the same magnetic order, suggesting that the two limits are adiabatically connected.

\section{Discussion and Summary}
\label{sec:discussion}

Recently, various measurements on Sr$_3$Ir$_2$O$_7$ using different techniques, such as ARPES\cite{Wang:arXiv1210.4141, Wojek:JPCM2012},
neutron scattering\cite{Dhital:100401}, and RIXS\cite{BJKIM1, BJKIM2, Fujiyama:2012}, 
in addition to transport\cite{Ge:2011yq,Cao:2002uq} and optical studies\cite{Moon:2008ly}, have been reported.
It is important to ask whether the current microscopic theory is compatible with the measured quantities, and further, if it offers
a useful starting point for future studies.

First of all, the latest ARPES data\cite{Wang:arXiv1210.4141, Wojek:JPCM2012} provide the dispersion of the occupied bands.
Using Slater-Koster hopping integrals for d-orbitals and considering the local distortion of the octahedra as well as a distance factor
that is exponentially decaying with distance, we obtained the hopping parameters up to next-nearest-neighbor.
Note that the only independent parameters left were the SOC strength and the tetragonal distortion.
Comparing the computed band structures and ARPES data, we show that the band below the Fermi energy is indeed mainly $J_{\text{eff}} = 1/2$, except near the $\Gamma$ point where there is a large contribution from $J_{\text{eff}} = 3/2$.
Overall, our tight-binding band structure fits well with first principle calculations\cite{Moon:2008ly} and the recent ARPES data\cite{Wang:arXiv1210.4141, Wojek:JPCM2012}. 

Furthermore, the current mean field study shows that the jump in the magnetic moment at the first order transition is weaker in the bilayer iridates than the single layer iridates. This is due to the band structure being nearly insulating, advancing our understanding of the smaller magnetic moment in bilayer iridates\cite{Fujiyama:2012}. 
As a consequence, the size of the charge gap is smaller in the bilayer system compared with the single layer one, agreeing with transport\cite{Ge:2011yq,Cao:2002uq} and optical conductivity\cite{Moon:2008ly} data.

A remaining puzzle is the giant magnon gap of over 90 meV reported in the RIXS data\cite{BJKIM2}. Our microscopic model contains terms
that break the spin rotation symmetry, a source for the magnon gap. Using a second order perturbation theory,
this microscopic model yields a spin-wave spectrum with a magnon gap of approximately $35$ meV, assuming that $t = 200$ meV and $U/t = 4$. 
The gap originates from the c-axis anisotropic exchange terms being different from the in-plane ones.
Since the tight binding parameters fit the ARPES data well, this discrepancy in the size of the magnon gap
implies that the second order perturbation assuming a large $U/t$ limit is not appropriate to estimate the magnon gap reported in the RIXS spectrum,
and one should add higher order terms in the perturbation theory.
In this intermediate-$U$ range, the spin susceptibility using the random phase approximation might be a more appropriate approach
than the semi-classical spin wave theory applied to the spin model obtained in the large-$U$ limit.  
Our microscopic model is a useful starting point for future study of the spin susceptibility for general
wavevector and frequency.

In summary, we build a tight binding model for Sr$_3$Ir$_2$O$_7$ and show that the non-interacting system is almost a band insulator
where the two unoccupied $J_{\text{eff}}=1/2$ bands are separated from the rest of occupied bands by a direct gap at every k-point.  
As the interaction strength increases, a band insulator to magnetic insulator transition occurs,
and a finite magnetic moment pushes the top/bottom of hole/electron bands further away from the Fermi level making the direct band gap bigger.
This is qualitatively different from the single layer Sr$_2$IrO$_4$, where the degeneracy of the bands crossing the Fermi level along $M-X$
is lifted turning it into a magnetic insulator.\cite{Watanabe:2010cr}
The ground state magnetic ordering pattern is sensitive to the lattice structure, with the staggered rotation of IrO$_6$ octahedra between adjacent layers
playing a crucial role in both developing a band insulator in the tight binding spectrum and determining the canted AF ordering pattern.
Sr$_3$Ir$_2$O$_7$ is a magnetic insulator with a small moment\cite{Fujiyama:2012}, and thus it is close to the transition.
However, the AF ordering pattern obtained from the spin model derived in the large $U$-limit\cite{JM_arxiv_2012_iridates}
is identical to that obtained from the current mean-field theory,
implying that this small moment insulator is likely smoothly connected to the Mott insulating regime as $U$ increases, where the charge gap persists above the magnetic ordering temperature.

\textit{Acknowledgement -}
This work was supported by NSERC of Canada.

\bibliography{BibSrIrO}

\end{document}